%% file: sz_pol_aniso_final_nbf.tex
\documentclass[useAMS,usenatbib]{mn2e}
\usepackage{graphicx,amssymb,epstopdf,moresize,caption,multirow}
\input{./defs/jnames}
\input{./defs/refs}

\input{./defs/vecs}

\input{./defs/struct}

\input{./defs/math}

\input{./defs/units}
\input{./defs/local}
\title[SZ polarization due to pressure anisotropy]{Polarization of Sunyaev-Zeldovich signal due to electron pressure anisotropy in galaxy clusters}
\author[Khabibullin et al.]{
I. Khabibullin$^{1,2}$\thanks{E-mail: ildar@mpa-garching.mpg.de}, 
S. Komarov$^{2}$, 
E. Churazov$^{1,2}$, 
A. Schekochihin$^{3,4}$\\
$^{1}$Max Planck Institute for Astrophysics, Karl-Schwarzschild-Strasse 1, 85741 Garching, Germany\\
$^{2}$Space Research Institute (IKI), Profsoyuznaya 84/32, Moscow 117997, Russia\\
$^{3}$The Rudolf Peierls Centre for Theoretical Physics, University of Oxford, 
1 Keble Road, Oxford OX1 3NP, United Kingdom\\
$^{4}$Merton College, Oxford OX1 4JD, United Kingdom
} 
\begin{document}
\maketitle
\begin{abstract}

We describe polarization of the Sunyaev-Zel'dovich (SZ) effect associated with 
electron pressure anisotropy likely present in the intracluster medium (ICM). 
The ICM is an astrophysical example of a weakly collisional plasma where 
the Larmor frequencies of charged particles greatly exceed their collision 
frequencies. This permits formation of pressure anisotropies, 
driven by evolving magnetic fields via adiabatic invariance, or by heat fluxes.
SZ polarization arises in the process of Compton scattering of the cosmic microwave 
background (CMB) photons off the thermal ICM electrons due to the difference in the characteristic 
thermal velocities of the electrons along two mutually orthogonal directions in the sky plane. The signal scales linearly with the optical depth of the region containing large-scale correlated anisotropy, and with the degree of anisotropy itself. It has the same spectral dependence as the 
polarization induced by cluster motion with respect to the CMB frame (kinematic SZ effect 
polarization), but can be distinguished by its spatial pattern. { For the illustrative case of a galaxy cluster with a cold front,  where electron transport is mediated by Coulomb collisions, we estimate the CMB polarization degree at the level of 10$^{-8}$ ($\sim 10$ nK). An increase of the effective electron collisionality due to plasma instabilities will reduce the effect.  Such polarization, therefore, may be an independent probe of the electron collisionality in the ICM, which is one of the key properties of a high-$\beta$ weakly collisional plasma from the point of view of both astrophysics and plasma theory.}

\end{abstract}
\section{Introduction}
\label{s:intro}

The cosmic microwave background (CMB) radiation in the direction of galaxy clusters 
is distorted due to Compton scattering of the CMB photons off the hot electrons of 
the intracluster medium (ICM), as first predicted by \cite{SZ1972}. These distortions 
have a characteristic spectral shape determined by the relative contributions of the 
thermal (tSZ, \citealt{SZ1972}) and kinematic, i.e., related to the bulk motion of a 
cluster (kSZ, \citealt{SZ1980}), effects. Both are now readily detected by space and 
ground-based millimetre and sub-mm observatories (e.g., \citealt{Hasselfield2013,Bleem2015,Planck2014s}). 

Importantly, the amplitude of the effect is proportional to the volume-integrated gas 
pressure of the ICM (which can be used as a proxy for the cluster's mass), {and its surface 
brightness does not depend on the distance to the cluster}, making the SZ signal an 
extremely valuable tool for both cosmological and ICM studies \citep{SZ1981,Rephaeli1995,
Birkinshaw1999,Carlstrom2002,Planck2014c}. Recently, a high-resolution ($\sim5$'') mapping of 
the SZ effect became available owing to the \textit{ALMA} observatory \citep{Kitayama2016}. 
It allows to study physical scales of $\sim 20 $ kpc at the distance of $ z \sim 0.25 $ 
{with the sensitivity $17\,\mu$Jy beam$^{-1}$ (at 92 GHz), or $\sim 100\, \mu$K at $5$'' full width at half maximum (\citealt{Kitayama2016}, see also \citealt{Young2015} for MUSTANG/GBT and \citealt{Adam2016} for IRAM/NIKA detections).}

Because Compton-scattered photons are linearly polarized, the SZ effect also has 
the potential to reveal itself in the polarization of the CMB radiation \citep{SZ1980,
ZS1980, SZ1981}. For the net polarization signal from a cluster not to cancel out 
after integration over the incident photon momenta, the presence of a quadrupole 
component in the CMB angular anisotropy (as seen by the electrons in the cluster) is 
needed \citep{ZS1980}. Quadrupole anisotropy can be inherent to the CMB radiation 
itself \citep{ZS1980}, or can be induced by the motion of the cluster with respect to 
the CMB, finite-optical-depth effects, or a combination of these \citep{ZS1980,SS1999,
Lavaux2004,Shimon2006}. Additionally, the local intensity distribution can be distorted 
by gravitational effects, e.g., the moving-gravitational-lens effect \citep{Gibilisco1997}.

In this paper, we predict yet another potential mechanism of generation of the SZ 
polarization by electron pressure anisotropies. Such anisotropies are 
typically produced in a plasma where the Coulomb collision frequencies of the 
charged particles are small compared to their Larmor frequencies (such a plasma 
is often called weakly collisional). This is, indeed, the case in the ICM: even 
for a seemingly small magnetic field ($\sim 1 \mathrm{\mu G}$) observed in galaxy 
clusters \citep{Feretti2012}, Larmor scales are separated from Coulomb mean free 
paths by many orders of magnitude. Pressure anisotropies can then be driven by 
evolving magnetic fields via adiabatic invariance, and by heat fluxes (see e.g. 
\citealt{Schekochihin2006} and references therein). The amplitude of these anisotropies is 
likely to be small: first, because plasma motions in the ICM are typically significantly 
subsonic; second, because even a low anisotropy quickly leads to the development of 
kinetic microinstabilities. These instabilities (firehose, mirror and, possibly, whistler modes) regulate the anisotropy level by particle scattering off magnetic perturbations. As a result, 
the anisotropy is kept at the low level of marginal stability (e.g., \citealt{Kunz2014,Riquelme2016,Santos2016}; 
see also \citealt{Kasper2002,Hellinger2006,Stverak2008,Bale2009,Chen2016} for direct observations 
in the solar wind{, where anisotropies reach $ \gtrsim 1$ level in low ($ \lesssim 1$) plasma beta regions}). Whether this picture is fully applicable to the intracluster 
plasma is rather uncertain, and observational techniques offering a peek into plasma 
microphysics by constraining the effective electron collisionality of the ICM are of interest. 

One of the most promising targets for such studies can be galaxy clusters containing 
shocks and cold fronts, i.e., sharp temperature and density discontinuities associated 
either with a cold subcluster moving in a host cluster or sloshing of relatively cool 
gas displaced from a cluster core (\citealt{Markevitch2000}; see \citealt{Markevitch2007,
Zuhone2016} for reviews). Field-line draping along a cold 
front interface is believed to be responsible for keeping these substructures from 
smearing out by thermal conduction and hydrodynamical instabilities (\citealt{Ettori2000,
Vikhlinin2001,Vikhlinin2002}, also see \citealt{Churazov2004} for an alternative). 
Along with heat fluxes, these evolving magnetic fields should also produce both ion and electron pressure 
anisotropies spatially ordered on a macroscopic scale of a cold front. Large-scale 
anisotropy can also be produced by compression of magnetic field at shocks that 
form ahead of supersonic cold fronts. Therefore, the polarization signal induced by 
electron anisotropy can survive after integration along the line of sight, as 
opposed to the polarization produced by random turbulent motions.  

In the previous work {(\cite{Komarov2016b}, Paper I hereafter)}, we demonstrated this for the polarization 
of cluster thermal bremsstrahlung X-ray emission. We used a trans-sonic cold front 
with a bow shock as a numerical model to study the total polarization resulting from  
compression, field-line stretching and heat fluxes. Here, we take advantage of the 
same numerical setup to predict the corresponding CMB polarization and to compare it 
with the expected CMB polarization in the direction of galaxy clusters induced by 
other effects. For our illustrative case, the amplitude of anisotropy-induced 
polarization turns out be at the same level ($\sim 10 $ nK) as polarization induced 
by the motion of the subcluster and by finite-optical-depth effects. However, 
different spatial or spectral shapes of the signals could facilitate the 
differentiation between them.
  
The paper is structured as follows. In Section \ref{s:aniso}, we describe the generation 
of electron pressure anisotropies in the ICM. In Section \ref{s:szpolarization}, we 
calculate the corresponding CMB polarization and compare it with polarization 
signals due to other effects. We present the predicted polarization signal for a 
simulated cold front in Section \ref{s:results}, and discuss the feasibility of 
its detection in Section \ref{s:discussion}. The conclusions are given in 
Section \ref{s:conclusions}.       

\section{Pressure anisotropies in the ICM}
\label{s:aniso}

~~~~~~The ICM is a hot ($T\sim 10 $ keV), tenuous ($n\sim10^{-3}$ cm$^{-3}$) weakly 
magnetized plasma. The typical magnetic-field strength, $B\sim$ a few $\mu$G \citep[see][]
{Carilli2002,Feretti2012}), corresponds to the ratio of thermal to magnetic-energy 
densities (plasma beta) $ \beta_{pl}=8\pi nT/B^2\sim 100$. This immediately implies 
that for all particle species $s$, the mean free paths $\lambda_s$ set by Coulomb 
collisions greatly exceed the corresponding Larmor radii  $\rho_s$ by orders of 
magnitude. For a fiducial set of the ICM parameters, one gets \citep{Spitzer1962,Sarazin1988}:
\beq
\label{eq:mfp}
\lambda_{s}=\lambda_{mfp} \approx 20 ~\mathrm{kpc}~\left(\frac{T}{8~\mathrm{keV}}\right)^{2}
~\left(\frac{n}{10^{-3}\mathrm{cm}^{-3}}\right)^{-1}
\eeq
for both protons and electrons, and
\beq
\rho_p \sim 4\times10^{-12} ~\mathrm{kpc} ~ \left(\frac{T}{8~\mathrm{keV}}\right)^{1/2}~
\left(\frac{B}{1~\mu\mathrm{G}}\right)^{-1},
\eeq
\beq
\rho_e \sim 10^{-13} ~\mathrm{kpc} ~ \left(\frac{T}{8~\mathrm{keV}}\right)^{1/2}~
\left(\frac{B}{1~\mu\mathrm{G}}\right)^{-1}
\eeq
for the proton and electron Larmor radii, respectively.

Therefore, Coulomb collisions are ineffective in keeping particle distributions 
isotropic, and they become gyrotropic, i.e., axially symmetric around the 
field lines. In this case, a difference between 
the parallel and perpendicular (to the magnetic-field line) pressure 
can develop. It is characterized by the degree of anisotropy
\beq
\Delta_s \equiv \frac{p_{\perp s}-p_{\parallel s}}{p_s},
\eeq 
where $p_{\perp s} $ and $p_{\parallel s} $ are the perpendicular and parallel 
pressure, respectively, while $ p_s=\frac{1}{3}p_{\parallel s}+\frac{2}{3}p_{\perp s} $ 
is the mean pressure, all for particle species $s$. Apart from adiabatic invariance, 
heat fluxes can provide an additional source of pressure anisotropy. 

The value of $\Delta_s  $ is set by the balance between the rate of isotropisation 
by particle scattering { (either by Coulomb collisions or scattering off magnetic perturbations produced by plasma instabilities)} and the rate at which anisotropy is driven by changing 
magnetic fields, changing particle density, and heat fluxes, as described by 
the so-called modified ({ by the inclusion of isotropic collisions and heat fluxes}) CGL equations 
\citep{Chew1956,Schekochihin2010}. From them, it follows that
\bea
\nonumber
\Delta_s = \frac{\pperps - \ppars}{p_s} &\approx& 
			\frac{1}{\nu_s} \bigg [ \frac{1}{B} \frac{d B}{d t}  
		   -\frac{2}{3}\frac{1}{n_s}\frac{d n_s}{d t} \\
 &&+ \frac{4 \bnabla\cdot (q_s \vc{b}) - 6 q_s \bnabla \cdot \vc{b}}
 						{15 p_s} \bigg ],
\label{eq:tot_anis}
\eea
where $\nu_s$ is the {effective} collision frequency, $ B $ the magnetic-field strength, $\vc{b}$ 
the magnetic-field unit vector, $ n_s $ is the number density of the particle specie $s$. 
The total parallel heat flux is $ q_s=q_{\perp s}+q_{\parallel s}/2=(5/6) q_{\parallel s} $, 
where $q_{\perp s}$ and $q_{\parallel s}$ are the parallel flux of the ``perpendicular 
internal energy'' and the parallel flux of the ``parallel internal energy'' respectively 
(\cite{Schekochihin2010}; {Paper I}). The first two terms on the right side of 
equation~(\ref{eq:tot_anis}) correspond to the conservation of the magnetic moment of 
a charged particle (the first adiabatic invariant) in a weakly collisional plasma with 
evolving magnetic fields, while the last term is the contribution of the parallel heat 
flux.  

The collision timescales in the ICM are still short enough compared to the timescales 
set by thermal conduction and fluid motions feeding the anisotropy for $\Delta_s$ 
to be low, and $\ppars - \pperps \ll \pperps \approx 
\ppars \approx p_s$. Note that the electron Coulomb collision frequency declines steeply as the velocity of a particle 
increases, $\nu_e(v)\propto v^{-3}$ (e.g., \citealt{Spitzer1962}). This means that one might 
expect a somewhat higher level of anisotropy for suprathermal electrons. However, the number 
of such electrons drops even more rapidly ($\propto v^2 e^{-m_ev^2/2T}$), so their influence 
on the amplitude of the effects under consideration stays small, causing an amplification by a 
factor of the order of unity at most. 


The contribution of different driving terms $\Delta_{B,n;s}$ (changing $B$ and $n$) and 
$\Delta_{T;s}$ (heat fluxes) to the total anisotropy can be estimated as (e.g. {Paper I}) 
\bea
\label{eq:DB}
\lefteqn{\Delta_{B,n;s} \sim \frac{u}{\vths} \frac{\lambda_s}{L_u},}\\
\label{eq:DT}
\lefteqn{\Delta_{T;s} \sim \frac{\lambda_s^2}{L_T L_u} \frac{\delta T_s}{T_s},}
\eea
where { parallel (with respect to the magnetic field)} fluid motions are characterized by velocity $u$ at parallel scale $L_u$, variations 
of $B$ at the parallel scale of the velocity field $L_B=L_u$, and parallel temperature 
gradient $\nabla_{\parallel} T_s \sim \delta T_s/L_T$ at scale $L_T$. The heat 
flux is given by $q_s = - \kappa_s\nabla_{\parallel} T_s$, 
where thermal conductivity $\kappa_s \sim n_s \vths \lambda_s$, $\lambda_s$ is the 
mean free path and $\vths$ the thermal speed. 


Under conditions typical for the ICM, the term associated with the magnetic-field 
changes is 
likely to dominate the ion anisotropy, while for electrons, the contribution of 
thermal conduction can be of the same order, depending on the properties of the flow 
and the relative orientation of magnetic-field lines and temperature gradients 
(see, e.g., \citealt{Komarov2014}). It is the electron pressure 
anisotropy that primarily determines the expected polarization of both the thermal 
bremsstrahlung and SZ signals, so in general both anisotropy-driving 
terms have to be taken into account.  

Substituting the parameters typical of a cold front, which is the main example analyzed in this 
paper, we have $u \sim \vthi$ (i.e., a nearly 
sonic flow) and $ L_u\sim R$, where  $ R $ is the cold front's radius of curvature. 
Therefore, the expected 
level of the anisotropy is $\Delta_{B,n;p} \sim \lambda / R$ for ions and $\Delta_{B,n;e} 
\sim 1/40\times\lambda/R$ for electrons. For a cold front of radius $ R\sim 200 $ kpc, using 
the Coulomb mean free path (\ref{eq:mfp}), 
this results in $\Delta_{B,n;p} \sim 0.1$ for ions and $\Delta_{B,n;p} \sim 2.5\times 10^{-3}$ 
for electrons. As noted above, the total electron anisotropy may also include a 
comparable heat-flux contribution: by analyzing our numerical simulation data, we calculate both 
driving terms in Section \ref{s:results}.  


It should be noted that the maximum total anisotropy $\Delta_e+\Delta_p$ is bound by the 
thresholds of the firehose (from below) and mirror (from above) instabilities, which 
rein the anisotropy at marginal stability \citep{Schekochihin2006,Schekochihin2010,Kunz2014,Sironi2015,Burgess2016,Riquelme2016}. 
In the case of small anisotropy, the limits are $-2/\beta_{pl} <\Delta_e+\Delta_i < 1/\beta_{pl}$
\citep[e.g.,][]{Kunz2014}. Due to a high $ \beta_{pl}\sim100$ in the ICM, the net anisotropy 
(dominated by ions, as estimated above) presumably leads to ubiquitous generation of magnetic perturbations by ion 
(firehose and mirror) kinetic instabilities in regions with fluid motions. In addition to 
regulating the total anisotropy by scattering ions, these perturbations are able to enhance 
the electron collisionality as well, e.g., by magnetic mirroring \citep{Komarov2016a}. However, 
in regions where magnetic-field lines are stretched by large-scale fluid motions, as at the 
interface of a cold front, $ \beta_{pl}$ may be reduced sufficiently to avoid formation of 
the instabilities. We showed this in Appendix A of {Paper I} by comparing the calculated 
total anisotropies with the ion kinetic instabilities' thresholds. This means, the reader should 
be aware of the fact that in those regions where the development of instabilities is predicted, 
the resulting SZ polarization signal could be reduced because of the higher effective electron 
collisionality. 
Electron instabilities, on the other hand, can also be triggered even if the electron anisotropy is 
rather small (see, e.g., \citealt{Riquelme2016,Riquelme2017}, { or in the case of a plasma with a temperature gradient (\cite{Levinson1992,Pistinner1998,Roberg2017}; Komarov et al., in prep.).}
In all that follows, we assume that the electron anisotropy is unaffected by various instabilities 
and, thus, estimate the upper limit on the SZ polarization signal induced by the anisotropy.


The gyrotropic pressure anisotropy is commonly described by means of the bi-Maxwellian velocity distribution function: 
\bea
\nonumber
f(v,\theta_0)&=&n_e\left(\frac{m_e}{2\pi T_{\perp}}\right)\left(\frac{m_e}{2\pi T_{\para}}\right)^{1/2}\\ 
	&&\times \exp\left[-\frac{m_e v^2}{2 T_0}\left(\frac{T_0}{T_{\perp}}\sin^2\theta_0+\frac{T_0}{T_{\para}}\cos^2\theta_0\right)\right],
\label{eq:fv}
\eea
where $v$ is absolute value of the velocity, $ \theta_0 $ is the angle between the velocity vector and the symmetry axis (i.e., the magnetic field direction), and $T_0=(1/3) T_{\para} + (2/3) T_{\perp}$ is the mean temperature.

In the case of small anisotropy $\Delta \equiv (T_{\perp}-T_{\para})/T_0$ we have $\Delta m_e v^2 / (2 T_0) \ll 1$ for the bulk of the electron population, so one can expand this distribution function to the first order in $\Delta$ as 
\beq
\label{eq:fvexp}
f(v,\theta_0)=f_0(v)+\delta f_\Delta(v,\theta_0),
\eeq
where $f_0(v)$ is an isotropic Maxwell distribution at temperature $ T_{0}$:
\beq
\label{eq:f0}
f_0(v)=n_e\left(\frac{m_e}{2\pi T_{0}}\right)^{3/2}\exp\left(-\frac{m_e v^2}{2 T_0}\right),
\eeq
while the anisotropic part of the distribution is 
\beq
\label{eq:df}
\delta f_\Delta(v,\theta_0)=\Delta ~\frac{m_e v^2}{2 T_0}~\left(\frac{1}{3}-\cos^2\theta_0\right) ~ f_0(v).
\eeq
Defining $ \mu=\cos\theta_0$, we rewrite this as 
\beq
\label{eq:dfleg}
\delta f_\Delta(v,\mu)=-\Delta ~\frac{m_e v^2}{2 T_0}~ f_0(v)~P_{2}(\mu),
\eeq
where $P_{2}(\mu)=\mu^2-1/3$ is the Legendre polynomial of the second order, so this perturbation is of purely quadruple nature, with the relative amplitude $ \delta f_\Delta(v,\mu)/f_0(v)$, however, rising towards larger $v$ as $ \propto v^2$.
As mentioned above, the actual anisotropy of the suprathermal electrons might turn out to be higher due to their lower collisionality compared to the bulk population ($\nu_s\propto v^{-3}$), so the relative anisotropy amplitude might rise as $ \propto v^{5}$. However, even if such scaling indeed takes place, it leads only to a factor of 1.3 increase in the predicted polarization signal (see Section \ref{s:szpolarization}).  


\section{SZ polarization}
\label{s:szpolarization}
\begin{figure}
\includegraphics[scale=1,width=0.95\columnwidth, bb= 60 210 550 720]{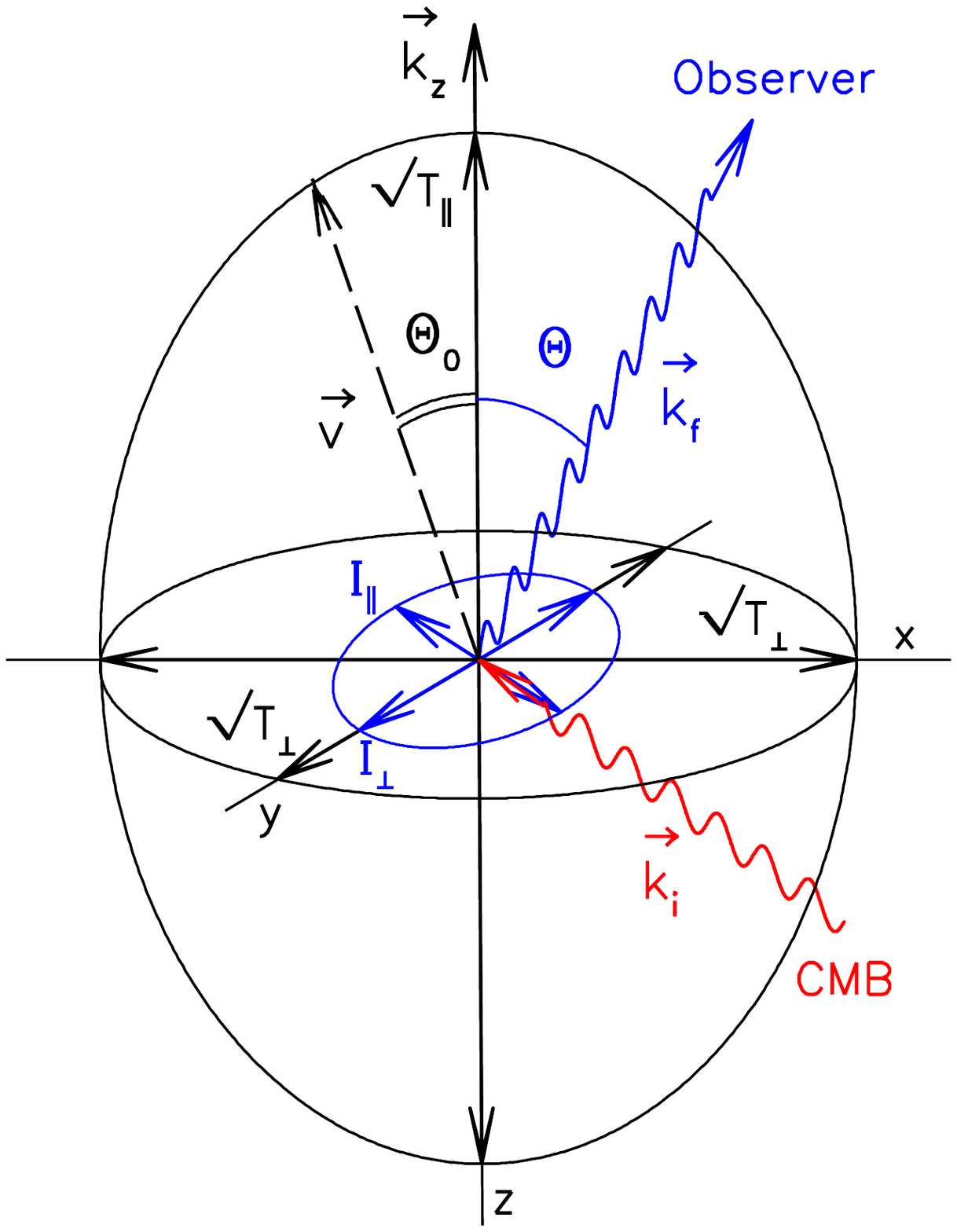}
\caption{Illustration of the geometry of the problem. The incident CMB photon's direction is shown as $\vc{k_i}$, the direction of the scattered photon is $\vc{k_f}$, $\vc{k_z}$ is aligned with the system's symmetry axis, set either by the CMB quadrupole axis, or the velocity direction of an electron, or the magnetic-field direction in the case of a gyrotropic velocity distribution. In the latter case, the ellipsoid depicts an isoprobability surface in the velocity space, with the major and minor axis proportional to $ \sqrt{T_{\parallel}}$ and $ \sqrt{T_{\perp}}$, respectively. The blue ellipse demonstrates the coordinate system in the picture plane of an observer, where $ I_{\parallel}$ and $ I_{\perp}$ correspond to photons with the electric-field vector oscillating along the projection of $\vc{k_z}$ and perpendicular to it, respectively.}
\label{f:sketch}
\end{figure}

Let us consider the CMB radiation field with the intensity given by
\beq
I_{\nu}(x)=C\frac{x^3}{e^x-1}
\label{eq:cmb}
\eeq  
where $x=h\nu/kT_{cmb}$, $ C=2(kT_{cmb})^3/(hc)^2$, $ T_{cmb}=2.725$ K, $h$ is the Planck constant and $c$ is the speed of light. 

As was shown already by \citealt{SZ1980}\footnote{In what follows, we will ignore relativistic corrections, which become important for gas temperatures above $ \gtrsim 10$ keV; see \cite{Challinor2000} and \cite{Itoh2000} for detailed discussions.}, the presence of a quadrupole component in this incident radiation field $ I_{\nu} $ (as seen by an electron) gives rise to linear polarization of the scattered CMB radiation with the polarization degree
\begin{equation}
P_{\nu}(\mu')=\frac{1}{10}(1-\mu'^2)\frac{I_2}{I_0}, 
\label{eq:polardeg}
\end{equation}
where $ I_0 $ and $ I_2 $ are the monopole and quadrupole amplitudes in the Legendre expansion of the incident radiation field $ I_\nu(\mu_0)=I_{0}+I_{1}\mu_0+I_{2}(\mu_0^2-1/3)+...$, and $\mu'=\cos\Theta=\vc{k_z}\cdot \vc{k_f}$ is the cosine of the angle between the quadrupole axis $ \vc{k_z} $ and the direction of the scattered photon $ \vc{k_f}$ (see Figure \ref{f:sketch}; \citealt{SZ1980}). The observed polarization vector will be aligned with the vector product of $ \vc{k_z} $ and $ \vc{k_f}$, so it will be perpendicular to the projection of the quadrupole axis on the picture plane. 

\subsection{Scattering of an intrinsic quadrupole}
\label{ss:quadrupol}

If there is an intrinsic quadrupole component in the angular power spectrum of the CMB corresponding to a temperature variance at level $ \delta T_{q}$, the predicted polarization signal in the direction of a galaxy cluster with the Thompson optical depth $ \tau\sim n_e L\sigma_T$ will be 
\beq
Q_{\nu,q}(x,\mu')=\frac{1}{10}\tau\frac{\delta T_{q}}{T_{cmb}}\varphi_{0}(x)I_{\nu}(x)(1-\mu'^2),
\label{eq:quadpol}
\eeq  
where $ \sigma_{T}=6.65\times 10^{-25}$ cm$^{2}$ is the Thomson scattering cross section, $ n_e $ characteristic electron number density and $ L $ the size of the cluster \citep{SS1999}.
In this relation, the function
\beq
\varphi_{0}(x)=\frac{d \ln I_\nu(x)}{d\ln T}=\frac{x e^x}{e^x-1}
\label{eq:phi}
\eeq
describes the spectral dependence of the polarization fraction $ P_{\nu}= Q_{\nu,q}(x)/I_{\nu}(x)$, while the factor $ 1-\mu'^2 $ originates from the amplitude dependence on the position of the cluster on the sky. 

The sky-averaged (rms) signal is 
\beq
Q_{\nu,rms}(x)=\frac{\sqrt{6}}{10}\frac{Q_{rms}}{T_{cmb}}\tau\varphi_{0}(x)I_{\nu}(x),
\label{eq:quadpolav}
\eeq  
where $ Q_{rms}$ is the rms amplitude of the quadrupole component. Since in the Local Universe it is measured (although with a large uncertainty) at the level of $ Q_{rms}\sim 10 \mu K$ (e.g., \citealt{Bennett2003,Bennett2013,Planck2014q}), the corresponding CMB polarization is expected at the level of $3\tau\mu K$ on average (while the maximum polarization is by a factor of $\approx 1.7$ higher; \citealt{SS1999}). 

Thus, for a galaxy cluster with $\tau \sim 10^{-2}$, $ Q_{rms}$ is expected at the level of $\sim25$ nK. Obviously, choosing not very distant galaxy clusters projected close to the direction of the CMB quadrupole axis should significantly decrease this effect, while its morphology could be easily predicted since it should closely follow the morphology of the cluster's optical depth.

\subsection{Scattering on a moving electron}
\label{ss:singleelectron}

Let us now consider an electron moving with respect to the thermal radiation background with velocity $v=\beta c$, where $ c $ is the speed of light. Assuming that the background radiation field is isotropic and has a blackbody spectrum with temperature $ T_{cmb} $, one can evaluate the spectral intensity of this radiation field as seen in the rest frame of the electron:
\begin{equation}
I_\nu(x,\mu_0) =C\frac{x^3}{e^{x\gamma_r(1+\beta\mu)}-1},
\label{eq:cmbinerf}
\end{equation}
where $\gamma_r=(1-\beta^2)^{-1/2}$ and $\mu_0$ is the cosine of the angle\footnote{As measured in the rest frame of the electron. In any event, transformation from the electron's rest frame to the CMB (and observer's) rest frame results in changes of order higher than second in $\beta$, so we will neglect them here.} between the electron's velocity vector and the direction of incidence of a photon \citep{SZ1980,SS1999}. 

Expanding equation~(\ref{eq:cmbinerf}) in Legendre polynomials and keeping terms up to second order in $\beta$ results in \citep{SS1999}  
\begin{eqnarray}
I_\nu(\mu_0)=C\frac{x^3}{e^{x\gamma}-1}\left[1+\frac{e^{x}(e^{x}+1)}{6(e^{x}-1)
^2}x^2\beta^2-
\right.
\nonumber\\
\left.
\frac{e^{x}}{e^{x}-1}x\beta\mu_0+\frac{e^{x}(e^{x}+1)}{2(e^{x}-1)^2}x^2\beta^2\left(\mu_0^2-\frac{1}{3}\right)+...\right],
\label{eq:cmbinerfexp}
\end{eqnarray}
so the amplitude of the quadrupole term is
\beq
I_2(x)=I_0(x) \varphi_{k}(x) \beta^2,
\label{eq:i2kin}
\eeq
with the quadrupole axis aligned with the electron's velocity direction, and the spectral dependence given by
\beq
\varphi_{k}(x)=\frac{e^{x}(e^{x}+1)}{2(e^{x}-1)^2}x^2.
\label{eq:fkin}
\eeq

Combining this with equation~(\ref{eq:polardeg}) gives
\beq
P_{\nu}(\mu')=\frac{1}{10}\varphi_{k}(x)\beta_t^2(1-\mu'^2), 
\label{eq:polardegkin}
\eeq
where $\beta_t=\beta\sqrt{1-\mu'^2}$ is the projection of the electron's velocity on the plane of the sky. The electric field of the polarized emission is perpendicular to the projection of the electron's velocity on the plane of the sky.

\subsection{Anisotropy-induced polarization}
\label{ss:aszpol}

The polarization of scattered emission considered above will vanish after the integration over isotropic electrons. If the distribution function is, however, not fully isotropic, some degree of polarization can be retained, and this is specifically the case for the gyrotropic distribution function considered in Section \ref{s:aniso}. 

Indeed, let the symmetry axis of the system (set by the local magnetic-field direction) be aligned with the $z$ axis, the $y$ axis be perpendicular to both this direction and the direction toward the observer and the $x$ axis lie in the same plane as the $z$ axis and the line of sight (see Figure \ref{f:sketch}). Being perpendicular to the line of sight, the $y$ axis lies in the picture plane, so we can use it as a reference axis for one of the Stokes parameters of the polarized emission, e.g., $Q$. Clearly, the other axis is then aligned with the projection of the $z$ axis on the picture plane. With such a choice of the coordinate system, the Stokes parameter $U$ should cancel out as a result of the axial symmetry of the system (the Stokes parameter $V$ is also zero because Compton scattering generates linear polarization only). 

Let us define $ I_{\perp}$ as the intensity of the scattered radiation with the electric-field vector oscillating perpendicular to the projection of the $z$ axis on the picture plane, i.e., along the $y$ axis, and $ I_{\parallel}$ analogously, but with with the electric-field vector oscillating parallel to it (see Figure \ref{f:sketch}).  According to the considerations in Section \ref{ss:singleelectron}, the contribution of an electron to $ I_{\parallel}$ is fully determined by the $y$ component of its velocity $ v_y=\beta_y c$:
\beq
P_{\nu,\parallel}=\frac{I_{\parallel}}{I_{0}}=\frac{1}{10}\varphi_{k}(x)\beta_y^2,
\eeq
while the contribution to $ I_{\perp}$ is determined by the projections of its $x$ and $z$ velocity components, $ v_x=\beta_x c$ and $ v_z=\beta_z c$ on the picture plane, viz.,
\beq
P_{\nu,\perp}=\frac{I_{\perp}}{I_{0}}=\frac{1}{10}\varphi_{k}(x)(\beta_x^2 \cos^2\Theta+\beta_z^2 \sin^2\Theta),
\eeq
where $ \Theta$ is the angle between the $z$ axis and the direction towards the observer (see Figure \ref{f:sketch}). 

Averaging these expressions over the axially symmetric electron distribution results in the substitution of $ \beta_x^2$, $ \beta_y^2$ and $ \beta_z^2$ by $ \left<\beta_\perp^2\right>$, $\left<\beta_\perp^2\right>$ and $\left<\beta_\parallel^2\right>$, respectively, these averages being proportional to $ T_\perp $ and $ T_\parallel$.  The net polarization $ P_{\nu,a}(\mu')=P_{\nu,\parallel}-P_{\nu,\perp} $ is then given by 
\beq
P_{\nu,a}(\mu')=\frac{1}{10}\varphi_{k}(x)(\left<\beta_{\perp}^2\right>-\left<\beta_{\parallel}^2\right>)\sin^2\Theta, 
\label{eq:polardegbimax}
\eeq
where the positive sign of $P_{\nu,a}(\mu')$ corresponds to the electric-field vector oscillating along the projection of the symmetry axis on the picture plane.

For a bi-Maxwellian electron distribution, one has $\left<\beta_\parallel^2\right>/(T_\parallel/m_e c^2)=\left<\beta_\perp^2\right>/(T_\perp/m_e c^2)=\eta\approx 1.3$, so the resulting polarization can be expressed as 
 \beq
P_{\nu,a}(\mu')=\frac{\eta}{10}\varphi_{k}(x)\Delta\frac{kT_0}{m_e c^2}(1-\mu'^2), 
\label{eq:polardegbimaxdelt}
\eeq
where we have replaced $\sin^2\Theta$ by $1-\mu'^2$ for similarity with the expressions~(\ref{eq:quadpol}) and (\ref{eq:polardegkin}) 
for other effects. Allowing for the existence of a power-law dependence of the anisotropy in the high-energy tails of the electron distribution (see Section \ref{s:aniso}) increases $\eta$ by a factor of $ 1.5$ at most. 

It is worth noting that only a large-scale observationally significant region of spatially correlated anisotropy in the ICM allows the combined polarization signal to be detectable. Any polarization produced by anisotropy fluctuations uncorrelated on scales much smaller than the size of a cluster (e.g., associated with turbulence) would be wiped out by integration along the line of sight. If the region of correlated anisotropies is characterized by the Thompson optical depth $ \tau_{a} $, then the corresponding net polarization signal is
\bea
\nonumber
Q_{\nu,a}(\mu')&=&P_{\nu,a}(\mu')\tau_{a}I_\nu(x)=\\
&&=\frac{\eta}{10}\Delta\frac{kT_0}{m_e c^2}(1-\mu'^2)\tau_{a}\varphi_{k}(x)I_\nu(x).
\label{eq:qaicl}
\eea

For mean electron temperature $ kT_0=0.01 m_e c^2=5.1$ keV and pressure anisotropy $ \Delta=10^{-3} $, the polarized signal is expected at the level $ \sim 35 (\tau_{a}/10^{-2})$ nK under geometrically most favourable conditions, i.e., when $\mu'=0$. Thus, the anisotropy-induced polarization might turn out to be of the same order as the polarization induced by scattering of the primary quadrupole, as well as the polarization produced by the other effects considered in what follows.

\subsection{Other sources of polarization}
\label{ss:other}
\begin{figure}
\includegraphics[width=\columnwidth, bb= 40 180 550 690]{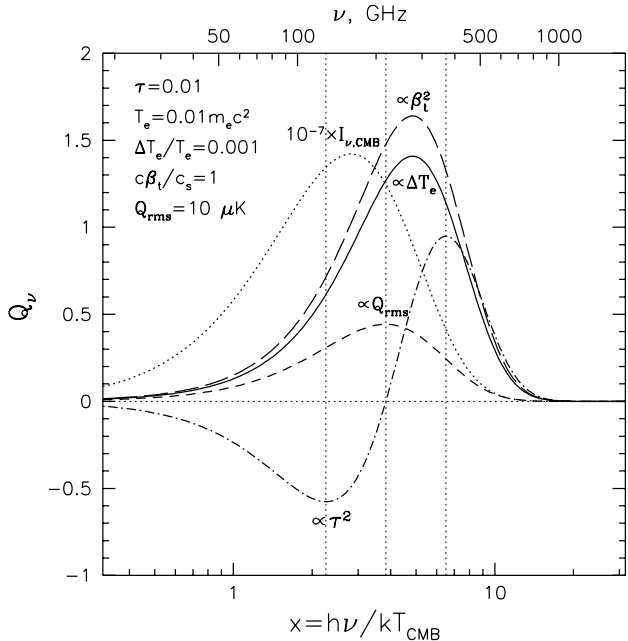}
\caption{Relative amplitudes and spectral dependences of various polarization signals compared to the CMB intensity multiplied by $ 10^{-7}$ (dotted line) for a cloud of electrons with temperature $ T_e=0.01m_ec^2=5.1$ keV, Thomson optical depth $\tau=0.01$, electron pressure anisotropy $ \Delta T_{e}/T_e=10^{-3}$, and moving in the direction perpendicular to the line of sight with velocity $ \beta_{t}c$ equal to the adiabatic speed of sound in the cloud $ c_s=\sqrt{\gamma T_e/\mu m_p} $, $\gamma=5/3$ and $\mu=0.6$, with respect to the CMB radiation field (characterized by quadrupole rms amplitude $ Q_{rms}=10 \mu K$) . The solid line shows polarization induced by pressure anisotropies, the long-dashed line by the kinematic SZ effect, the short-dashed line by scattering of the intrinsic CMB quadrupole, the dash-dotted line by second scatterings.}
\label{f:spec}
\end{figure}

In addition to the two effects considered above (viz., scattering of the CMB quadrupole and the presence of pressure anisotropies), there are a number of other effects capable of producing CMB polarization at approximately the same level. Namely, polarization arises due to the bulk motion of the cluster in the plane of the sky (kinematic SZ polarization) and due to scattering of the photons that have already scattered once off the electrons in the same cluster (called the $\tau^2$ polarization because it is a second-order effect in $\tau$). Besides that, polarization can be induced by the moving-gravitational-lens effect \citep{Birkinshaw1983,Birkinshaw1989,Gibilisco1997,Aghanim1998} and by the rotation of the cluster as a whole \citep{Chluba2002}, but the corresponding signal is likely to be 1-2 orders of magnitude smaller (see Table 
\ref{t:effects}), so we will not consider them here.    

\subsubsection{kSZ polarization}
\label{sss:kszpol}

Any bulk motions of the ICM (with respect to the CMB radiation field) that have a non-zero component in the plane of the sky also give rise to CMB polarization in that direction \citep{SZ1980}. For an ICM region (e.g., a subcluster) with optical depth $ \tau_{kin}$ moving with bulk transverse velocity $ \beta_t$, the polarization signal equals
\beq
Q_{\nu,k}(x)=\frac{1}{10}\beta_{t}^2\tau_{kin}\varphi_{k}(x)I_{\nu}(x),
\label{eq:qkin}
\eeq
as follows directly from equation~(\ref{eq:polardegkin}) \citep{SZ1980,SS1999}.

The characteristic scale of bulk motions is naturally set by the sound speed of the hot ICM $c_s=\sqrt{\gamma k T_0/\mu m_p}$, $\gamma=5/3$ and $\mu=0.6$, so one has $ \beta_t=M c_s/c $ with the Mach factor $M$ unlikely to exceed unity by a large factor (e.g., \citealt{Dolag2013}). Therefore, equation~(\ref{eq:qkin}) can be rewritten as 
\beq
Q_{\nu,k}(x)=\frac{1}{10}\frac{m_e}{m_p}\frac{\gamma}{\mu} M^2\frac{kT_0}{m_e c^2}\tau_{kin}\varphi_{k}(x)I_{\nu}(x).
\label{eq:qkinmach}
\eeq
Comparing this with equation~(\ref{eq:polardegbimaxdelt}) gives
\beq
\frac{Q_{\nu,k}(x)}{Q_{\nu,a}(x)}=\frac{m_e}{m_p}\frac{\gamma M^2}{\eta\mu\Delta}\frac{\tau_{a}}{\tau_{kin}},
\label{eq:qkinqanisorat}
\eeq
where $ \tau_{a} $ and $ \tau_{kin} $ are the characteristic Thompson optical depths of the regions with correlated electron pressure anisotropies and bulk motions, respectively. 

Clearly, for $ \Delta\sim 10^{-3}$ and $M=1$, one has $ Q_{\nu,k}(x)\sim Q_{\nu,a}(x) $ if $\tau_{a}\sim\tau_{kin}$, so the two effects are expected to be of the same order of magnitude, i.e., at the level of $ \sim 15(\tau_{k}/10^{-2})~nK$. This is particularly the case for galaxy clusters containing (super)sonic cold front substructures, as confirmed by the numerical simulations we present in Section \ref{s:results} (see also \citealt{Diego2003} for simulations of kSZ-induced polarization for a similar setup).

Bearing in mind identical spectral dependences of the two effects and the fact that the bulk kinematic motions responsible for the polarization signal are not probed by the kSZ spectral distortions, since the latter are determined (in the leading order) by the line-of-sight velocities, one has to rely on morphological separation of the signals aided with the X-ray/SZ mapping of hydro- and thermodynamic properties of the particular system under consideration. 

\subsubsection{$ \tau^2 $ polarization}
\label{sss:tauszpol}

Another source of CMB polarization in the direction of galaxy clusters arises from the fact that the CMB sky appears distorted for the electrons inside a galaxy cluster due to scatterings by other electrons of the same galaxy cluster, i.e., due to scattering of the primary thermal or kinematic SZ distortions \citep{SZ1980,SS1999,Lavaux2004,Shimon2006}. As a result, this effect is second-order in the cluster's optical depth $ \tau $. Spectral dependence of the polarization signal in this case corresponds to the spectral dependence of the primary SZ distortion, which for the thermal SZ is given by 
\beq
\varphi_{t}(x)=\frac{x e^{x}}{e^{x}-1}\left(x\frac{e^{x}+1}{e^{x}-1}-4\right),
\label{eq:fthermal}
\eeq
while for the kinematic SZ effect it is given by equation~(\ref{eq:phi}) \citep{SZ1972,SZ1980}.

For a homogeneous spherical cloud, the maximum intensity of the polarized emission is
\beq
Q_{\tau T, \nu}(x)=0.014 \frac{kT_0}{m_e c^2}\tau^2\varphi_{t}(x)I_{\nu}(x)
\label{eq:tauqtherm}
\eeq
for the thermal effect, and
\beq
Q_{\tau K, \nu}(x)=0.025 \sqrt{\frac{kT_0}{m_e c^2}}\sqrt{\frac{\gamma m_e}{m_p}}M\tau^2\varphi_{0}(x)I_{\nu}(x)
\label{eq:tauqkin}
\eeq
for the kinematic effect \citep{SZ1980,SS1999}. The ratio of these two effects is
\beq
\frac{Q_{\tau K, \nu}(x)}{Q_{\tau T, \nu}(x)}=M\frac{\sqrt{\gamma m_e/m_p}}{\sqrt{kT_0/m_e c^2}}\frac{\varphi_{0}}{\varphi_{t}}\approx 0.3M \sqrt{\frac{0.01}{kT_0/m_e c^2}}\frac{\varphi_{0}}{\varphi_{t}},
\label{eq:tauqratio}
\eeq
so the total $ \tau^2$ polarization is likely to be dominated by the scattering of the thermal SZ photons, except for frequencies close to $ x=3.83$, where the thermal effect changes sign. For a galaxy cluster with $ \tau\sim10^{-2}$, the corresponding polarization is expected at the level $ \sim40 (\tau/0.01)^2 ~nK$, however, it has a distinct spectral shape as compared with the effects considered above.

For a non-spherically symmetric cluster, the contribution of this effect is likely to be enhanced, so that its morphology reflects the relative distribution of matter inside the cluster \citep{Lavaux2004,Shimon2006}. In Section \ref{s:results}, we will calculate the expected signal for a galaxy cluster with a cold front following the approach outlined by equations (25) and (26) in \citep{SS1999}.

\section{Predictions for a cluster with a (super)sonic cold front}
\label{s:results}
\begin{figure*}
\includegraphics[scale=1,width=1.0\textwidth]{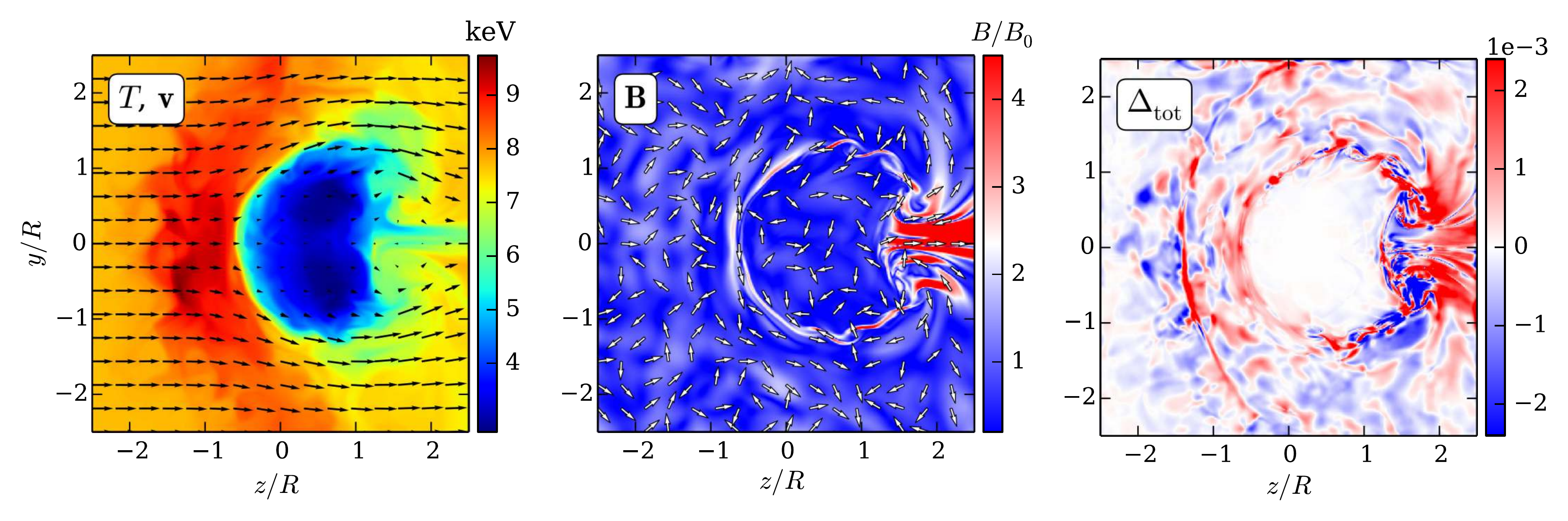}
\caption{An illustration of the 3D MHD simulation with a random magnetic field (correlation length $l_B\approx 100$ kpc) adapted from \citep{Komarov2016b}. All panels show a central slice of the computational domain (which is a 1 Mpc x 1 Mpc x 1 Mpc box) at the time $ t\approx 0.3$ Gyr after the start of the simulation. The left panel shows the temperature map (colour) and the velocity field (arrows). The magnetic field $\vc{B}$ is shown in the middle panel (colour: field strength; arrows: unit vectors in the magnetic field's direction projected onto the plane of the slice). The right panel shows the total generated pressure anisotropy defined relative to the local magnetic field's direction, where plus sign corresponds to the perpendicular pressure exceeding the parallel pressure.}
\label{f:simul}
\end{figure*}

As we have shown in Section \ref{s:szpolarization}, there are a number of effects that lead to polarization of the CMB in the direction of galaxy clusters, although the predicted morphologies and spectral dependences of the signals are different. This situation is very well illustrated by the example of a galaxy cluster containing a (super)sonic cold front \citep{Markevitch2000,Vikhlinin2001,Markevitch2007}. In this case, not only are significant bulk motions and asymmetries in the matter distribution present (which give rise to the kSZ and $\tau^2$ polarization signals; see \citealt{Diego2003}), but also magnetic-field stretching and sharp temperature gradients can occur, which potentially seed electron pressure anisotropies ({Paper I}). In this section, we take advantage of the 3D MHD simulation of such a system presented in {Paper I} in order to predict qualitatively the amplitude and morphology of the anisotropy-induced CMB polarization in comparison with other polarization sources.        

\subsection{Numerical setup}
\label{ss:simulation}

Our numerical setup consists of a 3D box region of hot dilute plasma ($T_{\rm out}=8$ keV, $n_{\rm out} = 10^{-3}$ cm$^{-3}$) of spatial extent $L=1$ 
Mpc containing a colder spherical subcluster ($T_{\rm in}=4$ keV) of radius $R=200$ kpc at its center. {The minimum linear scale captured by the simulation is $\approx$ 2 kpc. We use the same setup as in {Paper I}.}

Initial density distribution inside this radius is given by a beta model,
\beq
n_{\rm in} = n_c[1+(r/r_c)^2]^{-3\beta'/2},
\eeq
with $\beta'=2/3$, core radius $r_c=R/\sqrt{3}\approx 115$ kpc, 
and central density $n_c=8n_{\rm out}$.
The pressure balance is initially sustained by adding a gravitational acceleration field $\vc{g}$ to mimic the effect of a static dark-matter halo at the center of the computational domain. The simulation is run in the reference frame where the subcluster is initially at rest, so it starts with the surrounding hot gas uniformly overflowing the subcluster with velocity $v_0$, which is set to the sound speed in the hot ambient plasma, $c_{s0}=(\gamma 
p_{\rm out}/\rho_{\rm out})^{1/2}=(\gamma_{\rm gas} k T_{\rm out} 
/ \mu m_p)^{1/2} \approx 1400$ km/s, $\gamma=5/3$ and $\mu=0.6$. Such a setup is similar to the one considered by \cite{Diego2003}, who focused on the kSZ polarization produced in this situation, and \cite{Asai2007}, whose main focus was on stretching of the magnetic-field lines at the cold front interface.

The evolution of the system is calculated by solving a standard set of MHD equations with anisotropic thermal conduction.  { As we aim at setting an upper limit on the CMB polarization, we use the unsuppressed Spitzer thermal conductivity in all our runs. In order to calculate the pressure anisotropy associated with thermal conduction, electron heat fluxes are obtained self-consistently by calculating temperature gradients along the magnetic field and multiplying them by the Spitzer thermal conductivity, which is a strong function of temperature.}
Note that at large temperature gradients, heat flux may become saturated when the characteristic { parallel} scale of the gradients becomes comparable with the Coulomb mean free path. However, by calculating the mean free path over the computational domain, we have found that it is sufficiently small practically everywhere (even at the cold-front interface due to magnetic-field draping), so that saturation should not play a noticeable role in the evolution of temperature. We therefore do not modify our expression for the heat flux to include saturation. 
 
The initial magnetic field is set to be either uniform or random Gaussian with correlation length $ l_{B}=100$ kpc, and its strength corresponds to the plasma beta $ \beta_{pl}=200$ (in the case of a random field it is calculated with respect to the field dispersion $B_0^2=\left\langle B^2\right\rangle$). We stress that the final magnetic field in the simulation with a random initial distribution does have non-Gaussian statistics due to the fact that it is a product of dynamical evolution in the velocity field of the cold front. The uniform-magnetic-field case is invoked mainly to highlight the regions where the largest pressure anisotropies can potentially arise in such a system. The run with a random magnetic field illustrates how this idealized picture would change in the more realistic situation of a turbulent ICM where magnetic fields are tangled by random fluid motions.  The magnetic-field correlation length measured in the ICM is likely significantly shorter (see, e.g., \citealt{Vogt2005}). However, we believe that already for $l_B\lesssim R$, when field loops can be folded up against the cold-front interface, our model is capable of capturing the qualitative changes in the predicted picture compared to the uniform field case. We expect to see a factor of $\sim 2$ smaller polarization signal at the front in this case, because the random field lines are not oriented mainly perpendicular to the line of sight any more.  

The resulting flow and magnetic-field structures are shown shown in Figure \ref{f:simul} for one central slice through the computational domain, where the three main features are clearly visible: a (weak) bow shock ahead of the subcluster, the col- front interface, and the wake and backflow region that trails the main body of the subcluster. The generated pressure anisotropies are calculated according to equation~(\ref{eq:tot_anis}), and the resulting distribution of anisotropies (for random magnetic field) is illustrated by the rightmost panel of Figure \ref{f:simul}.  The highest level of anisotropy is predicted in the wake region, where turbulent vortices amplify the magnetic field by stretching along the direction of subcluster's motion (\citealt{Asai2007}; {Paper I}).     

\subsection{Polarization calculation}
\label{ss:polarcalc}

The output of the simulations described above is then post-processed to obtain the predictions for corresponding CMB polarization signals, which are calculated locally for each effect in terms of the Stokes parameter $dQ$ (according to expressions in Section \ref{s:szpolarization}), and then integrated along the line of sight (over the extent of the computational domain) by the standard procedure (e.g., \citealt{Lavaux2004}):
\beq
\label{eq:polarintergQ}
\tilde{Q}_{\nu,a}=\int d\tau \frac{dQ_{\nu,a}}{d\tau} \cos(2\chi),
\eeq
\beq
\label{eq:polarintergU}
\tilde{U}_{\nu,a}=\int d\tau \frac{dQ_{\nu,a}}{d\tau} \sin(2\chi),
\eeq
where $\chi$ is the angle between the local polarization axis (see Section \ref{s:szpolarization}) and the reference polarization axis chosen globally on the observer's picture plane (see, e.g., Figure 2 in {Paper I}).  

For our calculation of the kSZ polarization, we assume that the hot gas overflowing the cold clump is actually part of a bigger cluster that is at rest with respect to the CMB radiation field. As a result, the unperturbed outer gas does not contribute to the kSZ polarization, and the latter is fully dominated by the subcluster's contribution. 

To calculate the $ \tau^2 $ polarization, we first calculate it for a uniform box of a hot gas, with density and temperature equal to $ n_{out} $ and $T_{out}$ respectively, and then subtract it from the $\tau^2$ polarization calculated for the actual cold front setup. Both calculations are performed by evaluation of the CMB sky anisotropy due to the primary thermal SZ effect as seen by an electron at each particular point along the line of sight, and then integrating the resulting signal over the computational domain in the way similar to equations~(\ref{eq:polarintergQ}) and (\ref{eq:polarintergU}) (see \citealt{SS1999} for details).   

It is worth mentioning that the predictions for the kSZ and $\tau^2$ polarization differ only slightly between the uniform- and random-magnetic-field cases because the magnetic fields do not have a strong impact on the overall structure of the gas flow. 

\subsection{Predicted signal}
\label{ss:signal}

As was shown in Section \ref{s:szpolarization}, various sources of polarization are characterized by different spectral dependences, so we produced maps of the predicted signal in three spectral bands: $x=2.26$ (128 GHz), $x=3.83$ (218 GHz) and $x=6.51$ (370 GHz), at which the thermal SZ effect (and hence $\tau^2$ contribution) has maximum decrement, changes sign and has maximum increment, respectively (see dotted vertical lines in Figure \ref{f:spec}). The resulting spectrally resolved maps are shown in Figure \ref{f:szimage} for partial contributions from various effects and in Figure \ref{f:total} for the total predicted signal.

These results are consistent with the simple estimates presented in Section \ref{s:szpolarization}. The predictions for kSZ and $\tau^2$ polarization are also in line with the results of previous numerical simulations (e.g., \citealt{Diego2003,Lavaux2004,Shimon2006}). The partial contributions of these two effects are predicted at the level of $ \sim 10$ nK (see Figure \ref{f:szimage}), the total signal at the level of a few tens nK (see Figure  \ref{f:total}). Notably, the amplitude of the anisotropy-induced polarization indeed turns out to be comparable with the amplitude of the kSZ and $\tau^2$ polarization. However, these individual effects have either distinct morphologies of the signal in the observer's plane (e.g., kSZ polarization comes mainly from the gas of the subcluster, where anisotropy-induced polarization is negligible due to {higher collisionality of the cold and dense plasma there}), or distinct spectral shapes ($\tau^2$ polarization vanishes at $ x=3.83$), so one can hope to disentangle this complicated picture with the aid of high-angular-resolution and multi-frequency observations. 
      
\section{Discussion}
\label{s:discussion}
\input{./table1xx.tex}

The results of the simulations presented in Section \ref{s:results} confirm the basic predictions of Section \ref{s:szpolarization}: the electron pressure anisotropies potentially arising in the vicinity of a (super) sonically moving subcluster are capable of producing CMB polarization at the level comparable with the kSZ and $\tau^2$-induced signals. 

\subsection{Separating different sources of CMB polarization}

This level corresponds to $ \sim 10$ nK in the Rayleigh-Jeans part of the spectrum, and it is also comparable to the expected level of polarization induced by scattering of the CMB's intrinsic quadrupole (see Section \ref{s:szpolarization}). The contribution of the latter effect might be additionally suppressed by selecting the only clusters that lie in the direction of the local CMB quadrupole axis. Also, using X-ray and tSZ observations of the same cluster, one can (with a certain accuracy) reconstruct the density (and temperature) distribution inside the cluster and then exploit it to predict the corresponding CMB polarization signal. The same is true for the $\tau^2$ polarization, which can also be predicted based on the X-ray and tSZ maps of the cluster. Additionally, these two effects have spectral dependences different from kSZ and pressure-anisotropy-induced polarizations, so one can take advantage of multi-frequency observations to filter them out.

Separating the contribution of kSZ polarization appears to be the most challenging, since it has the same spectral dependence as the pressure-anisotropy-induced polarization. Also, it cannot be readily predicted from the observational data because it is determined by the difficult-to-measure transverse motions of the ICM plasma. In the case of a moderately supersonic motion of the subcluster, one may infer its velocity from the density jump measured by the X-ray surface brightness mapping, although such an estimate is likely to be prone to projection and line-of-sight averaging effects (see, e.g., \citealt{Markevitch2000,Vikhlinin2001}). Besides that, by making this estimate one measures the subcluster's velocity relative to its ambient ICM, and not with respect to the CMB radiation field, and the latter (which is actually what needed) can differ from the former due to the (likely unknown) peculiar transverse motion of the cluster as a whole. Still, as we have shown in Section \ref{s:results}, the predicted morphologies of the kSZ and anisotropy-induced polarizations are significantly different, with the latter being almost absent in the direction of colder and denser plasma. Thus, high angular resolution observations will be helpful to separate these two effects, complemented by an adequately fine and sensitive X-ray surface brightness and temperature mapping. {However, it is worth mentioning that high angular resolution is needed primarily for signal separation, rather than for detection of individual small-scale structures. Hence, the required sensitivity will be determined essentially by the signal integrated over regions of the correlated anisotropies. }  

Given that the collision rate goes down with the temperature, and, therefore, the anisotropy goes up, the anisotropy-induced polarization is expected to be significantly higher (at least as $\propto \lambda_{mfp}T_e\propto T_e^3$ for fixed $n_e$) in galaxy clusters that are more massive (and hotter) than the illustrative case considered here. Polarization signals due to other effects should also be higher in this case (because both optical depth and characteristic velocity of infalling subclusters should get higher as well), but their increase is likely to be less dramatic (e.g., as $\propto \beta_t^2 \propto T_e$ for kSZ-induced polarization). 

In addition to the CMB polarization caused by a galaxy cluster as described above, there are also primary fluctuations in CMB polarization at the angular scales of interest here ($\sim 1$ arcmin, i.e., $l\sim10^4$). These are significantly enhanced by gravitational lensing on the same cluster (or any intervening matter along the line of sight; see, e.g., \citealt{Lewis2006}). The expected amplitude of these fluctuations can be predicted based on our current knowledge of the CMB polarization power spectrum at larger scales and a mass model of the cluster, as has been extensively discussed in the literature (see, e.g., \citealt{Amblard2005,Liu2005,Maturi2007,Shimon2009,Ramos2012} for comparison of the level of these fluctuations with kSZ-induced polarization).

Secondary CMB polarization fluctuations generated during the re-ionization epoch should also contribute to the noise at these scales, but their level is likely to be orders of magnitude smaller \citep{Hu2000,Valageas2001,Santos2003,Zahn2005}. Potential contribution of point sources, e.g., lensed submillimetre galaxies (e.g., \citealt{Lima2010}), is even harder to predict, but it can potentially be tackled by deep observations at wavelengths that are capable of revealing possible counterpart sources.        

There are also a number of other possible sources of contamination, e.g., related to enhanced synchrotron emission due to the presence of a bow shock or scattering of radio emission of an AGN in one of the cluster's galaxies. This emission, however, typically has a spectral shape that is very distinct from the CMB and is more easily observed at significantly lower frequencies ($\lesssim 10$ GHz). Furthermore, there is likely to be contamination due to foreground polarized emission from the Galactic dust (e.g., \citealt{Tucci2005}), but this topic is far beyond the scope of our consideration here.

Certainly, the most promising technique would be to combine CMB polarization measurements with the measurements of polarization of X-ray emission both in lines (mainly from H- and He-like ions of heavy elements) and continuum (i.e., thermal bremsstrahlung emission), which should be sensitive to the same electron pressure anisotropies that determine the CMB polarization signal discussed here ({Paper I}). Besides that, pressure anisotropies in the distribution of the line-emitting ions might increase the broadening of these lines with respect to the broadening due to thermal and turbulent motions and resonant scattering. However, the collisionality of ions with charge $Z$ is boosted by a factor $ \sim Z^{1.5}$ compared to protons, so the expected amplitude of pressure anisotropies is comparable (e.g., for silicon and sulphur, $Z=14-16$) or smaller (for iron, $Z=25-26$) then the amplitude of electron pressure anisotropies, i.e., $ \sim 10^{-3}$. Although both of these techniques are out of reach of current and forthcoming X-ray facilities, further improvements in polarimetric and calorimetric technologies will finally make such studies feasible, improving the possibility to study pressure anisotropy-induced CM polarization as well.   

\subsection{Effective electron collisionality}

{ Detecting or constraining the CMB polarization associated with electron pressure anisotropies may allow one to extract valuable information about the microphysics of a high-$\beta$ plasma. If one knows the structure of the plasma flow (say, from current X-ray observations of cold fronts, or from the future precise X-ray measurements of gas velocities via ion lines), and thus the rate of change of magnetic fields,  it is possible to set lower limits on the electron collisionality. It is often believed that electron transport in the ICM is suppressed based on observations of significant temperature gradients in X-rays (e.g. \citealt{Markevitch2000,Ettori2000,Vikhlinin2001,Vikhlinin2002,Markevitch2007}). The exact mechanism of such suppression is yet to be understood, as it requires understanding of the intricate physics of a turbulent weakly collisional plasma. 

The mirror instability caused by turbulent stretching of magnetic-field lines has been identified as one of the possible suppression mechanisms. 
However, the typical suppression factors in the ICM unlikely exceed a factor of 1/5 \citep{Komarov2016a}. In Appendix A of {Paper I}, we marked the regions where the mirror instability can be triggered by the plasma flow past a cold front due to generation of positive pressure anisotropies.

Another interesting suppression mechanism is a whistler instability triggered by a heat flux. It can be shown that in a weakly collisional high-$\beta$ plasma, even a small heat flux goes unstable and inhibits itself by triggering slowly propagating transverse magnetic perturbations (\cite{Levinson1992,Pistinner1998,Roberg2017}; Komarov et al., in prep.). Nevertheless, large suppression factors are achieved only when the parallel temperature gradient scale approaches the electron mean free path, meaning for Knudsen numbers $\gtrsim 0.1$. Even in the case of cold fronts, where temperature gradients are the largest found in the ICM, parallel gradients turn out to be much smaller due to draping of the magnetic-field lines along the cold-front interface. We therefore conclude that currently, our understanding of the microphysics of a high-$\beta$ plasma does not allow one to predict very large suppression of electron transport confidently. 

Conversely, if, somehow, one knows that electron transport is mediated by Coulomb collisions in a certain region of a cluster, it is possible to estimate the parallel velocity shear associated with the magnetic-field stretching rate.}

\section{Conclusions}
\label{s:conclusions}

We have predicted the CMB polarization in the direction of a galaxy cluster containing a (super)sonic cold front due to electron pressure anisotropies potentially generated by magnetic-field-line stretching and heat fluxes in the weakly collisional plasma of the ICM. The amplitude of this signal is predicted at the level of $\sim 10$ nK in the Rayleigh-Jeans part of the spectrum, and it turns out to be comparable to the amplitude of kSZ and $\tau^2$ polarization, as well as of the polarization due to intrinsic CMB quadrupole scattering for such a cluster. 

With the aid of 3D MHD simulations {of a cold front}, we have demonstrated that the individual polarization effects have either distinct morphologies (e.g., the kSZ polarization comes mainly from the gas of the denser and colder subcluster, inside of which the anisotropy-induced polarization is negligible due to relatively higher effective collisionality of the gas), or distinct spectral shapes (e.g., the $\tau^2$ polarization vanishes at $ x=3.83$). As a result, one can hope to disentangle the resulting complicated picture by taking advantage of high-angular resolution and multi-frequency observations complemented by X-ray and spectral SZ data.

Measuring CMB polarization below 1 $\mu K$ currently presents an observational challenge. However, there is ongoing progress in available observational facilities, and also in our knowledge about various sources of contamination, and so accurately characterizing them is becoming more feasible. With further improvements, a detection of (or even an upper limit on) the pressure-anisotropy-induced polarization might become possible. This would provide a unique probe of the effective collisionality {(as well as the effective thermal conductivity) of the ICM} (paralleled perhaps only by X-ray polarization measurements, {Paper I}). { More fundamentally,} this will allow us to test the current understanding of the intricate physical processes animating `microscopic' scales in the hot turbulent weakly collisional plasma of the ICM. 

\section*{Acknowledgements}
We are grateful to the referee, Mark Birkinshaw, for the valuable suggestions that helped to  improve the paper noticeably. IK, SK and EC acknowledge partial support by grant No. 14-22-00271 from the Russian Scientific Foundation. AAS was supported in part by grants from UK EPSRC and STFC.

\newcommand{\labelstyle}{\Large}
\input{figure1}
\input{figure2}

\end{document}

%% file: defs/refs.tex
\newcommand{\eqref}[1]{equation~(\ref{#1})}

%% file: defs/vecs.tex
\newcommand{\vc}[1]{\bmath{#1}}

%% file: defs/struct.tex
\newcommand{\beq}{\begin{equation}}
\newcommand{\eeq}{\end{equation}}
\newcommand{\bea}{\begin{eqnarray}}
\newcommand{\eea}{\end{eqnarray}}

%% file: defs/math.tex
\newcommand{\bnabla}{\vc {\nabla}}

%% file: defs/local.tex
\newcommand{\vths}{v_{\rmn{th},s}}
\newcommand{\vthi}{v_{\rmn{th},i}}

\newcommand{\para}{\parallel}

\newcommand{\ppars}{p_{\parallel s}}
\newcommand{\pperps}{p_{\perp s}}

%% file: table1xx.tex
\begin{table*}
\caption{Summary of various sources of CMB polarization in the direction of galaxy clusters. The polarization degree is expressed as $P=P_0\alpha(T,\tau,\beta_t,...)\varphi(x)$, where $ P_{0}$ is the amplitude calculated for a fiducial set of parameters (see below), $\alpha(T,\tau,\beta_t,...)$ the scaling of the amplitude with these parameters, and  $\varphi(x)$, $x=h\nu/kT_{cmb}$, describes the spectral dependence of the signal. The fiducial set of parameters is $\tau=0.01$, $ kT_e=0.01m_ec^2=5.1$ keV, $ \Delta T_e/T_e=10^{-3}$, $ Q_{rms}=10~\mu K$,  $ \beta_tc=1000$ km/s, $\beta_rc=100$ km/s, $\Delta \theta=1$ arcmin, $D^{EE}_l=0.1~\mu K$ at $l=10^4$. Here, $Q_{rms}$ is the rms amplitude of the local CMB quadrupole component (e.g., \citealt{Bennett2003}), $ \beta_tc$ the cluster's transverse bulk velocity, $\beta_rc$ the circular velocity due to rotation of the cluster \citep{Chluba2002}, $\Delta\theta=4G M_{cl}/c^2R\approx0.7$ arcmin $ \left(\frac{M_{cl}}{10^{15}M_{\odot}}\right)\left(\frac{1~{\rm Mpc}}{R}\right)$  the angle of gravitational deflection of CMB photons by a cluster of mass $ M_{cl}$ at impact parameter $R$ \citep{Gibilisco1997}, $D^{EE}_l=l(l+1)C^{EE}_l/2\pi$ the E-polarization power spectrum amplitude at $l\sim10^4$ \citep{Lewis2006}.}
\begin{tabular}{lllll}
\hline\hline
Effect causing & Fiducial degree & \multirow{2}{*}{Scaling} & Spectral  & \multirow{2}{*}{Reference}\\
polarization &  of polarization  &  & dependence  & \\
\hline
\smallskip\smallskip
Pressure anisotropy & $\sim 10^{-8}$ & $\propto \frac{\Delta T_{e}}{T_e}\frac{kT_e}{m_e c^2}~\tau$  & $\frac{e^{x}(e^{x}+1)}{2(e^{x}-1)^2}x^2 $ & Section \ref{ss:aszpol}\\
\smallskip
CMB quadrupole & $\sim 10^{-8}$& $\propto \frac{Q_{rms}}{T_{cmb}}~\tau$  & $\frac{x e^x}{e^x-1}$ &Section \ref{ss:quadrupol}\\
\smallskip
Bulk motion (kSZ) &  $\sim 10^{-8}$ &$\propto \beta_{t}^2~\tau$ & $\frac{e^{x}(e^{x}+1)}{2(e^{x}-1)^2}x^2$ &Section \ref{sss:kszpol}\\
\smallskip\smallskip
Second scatterings ($\tau^2$) & $\sim 10^{-8}$& $\propto \frac{kT_{e}}{m_e c^2}~\tau^2$  & $\frac{x e^{x}}{e^{x}-1}\left(x\frac{e^{x}+1}{e^{x}-1}-4\right)$ &Section \ref{sss:tauszpol}\\
\smallskip
Moving lens & $\sim 10^{-9}$& $\propto \beta_{t}\Delta\theta~\tau $  & $\frac{x e^x}{e^x-1}$&\citealt{Gibilisco1997}\\
\smallskip
Cluster rotation & $\sim 10^{-10}$& $\propto \beta_{r}^2 ~\tau $  & $\frac{e^{x}(e^{x}+1)}{2(e^{x}-1)^2}x^2$& \citealt{Chluba2002}\\
\smallskip
CMB fluctuations & $\sim 10^{-8}$& $\propto \frac{\sqrt{D^{EE}_{l}}}{T_{cmb}}$  & $\frac{x e^x}{e^x-1}$ &\citealt{Lewis2006}\\
\hline
\end{tabular}
\label{t:effects}
\end{table*}

%% file: figure1.tex
\begin{figure*}
\centerline{\labelstyle  Kinematically-induced (kSZ) polarization}
\includegraphics[width=0.9\textwidth]{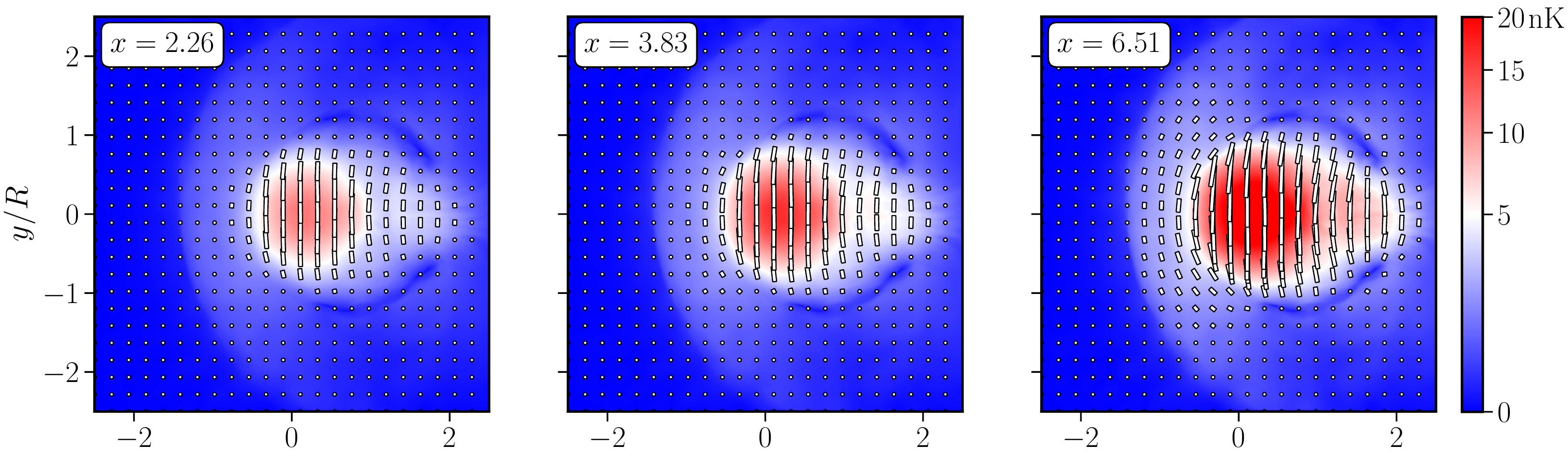}
\centerline{\labelstyle Second-scatterings-induced ($\tau^2$) polarization}
\includegraphics[width=0.9\textwidth]{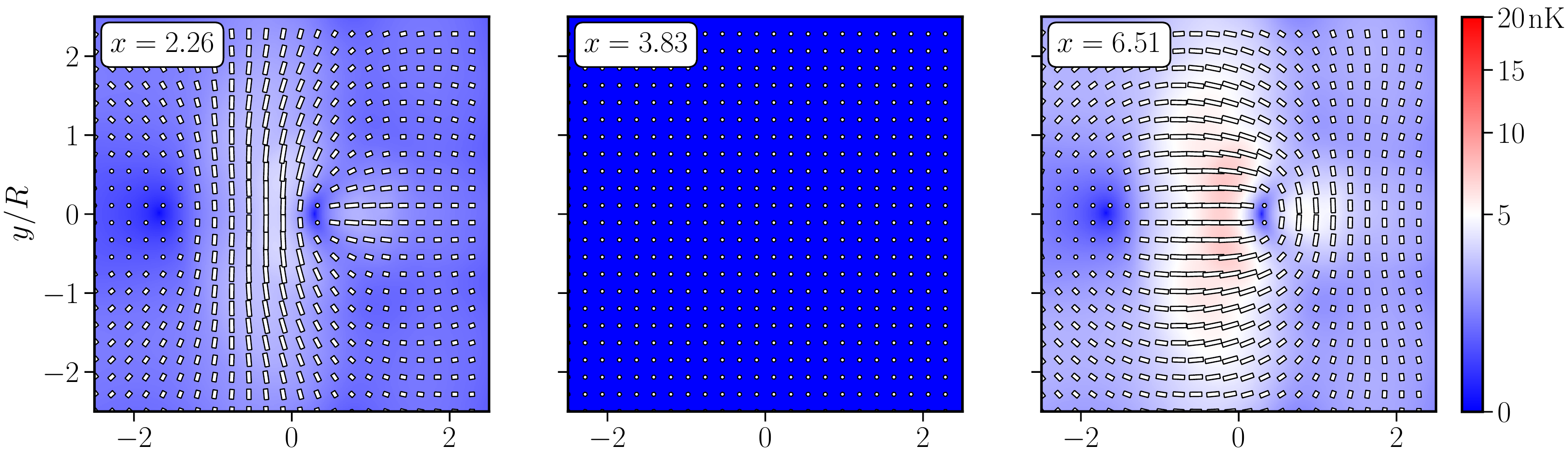}
\centerline{\labelstyle  Pressure-anisotropy-induced polarization, uniform initial magnetic field}
\includegraphics[width=0.9\textwidth]{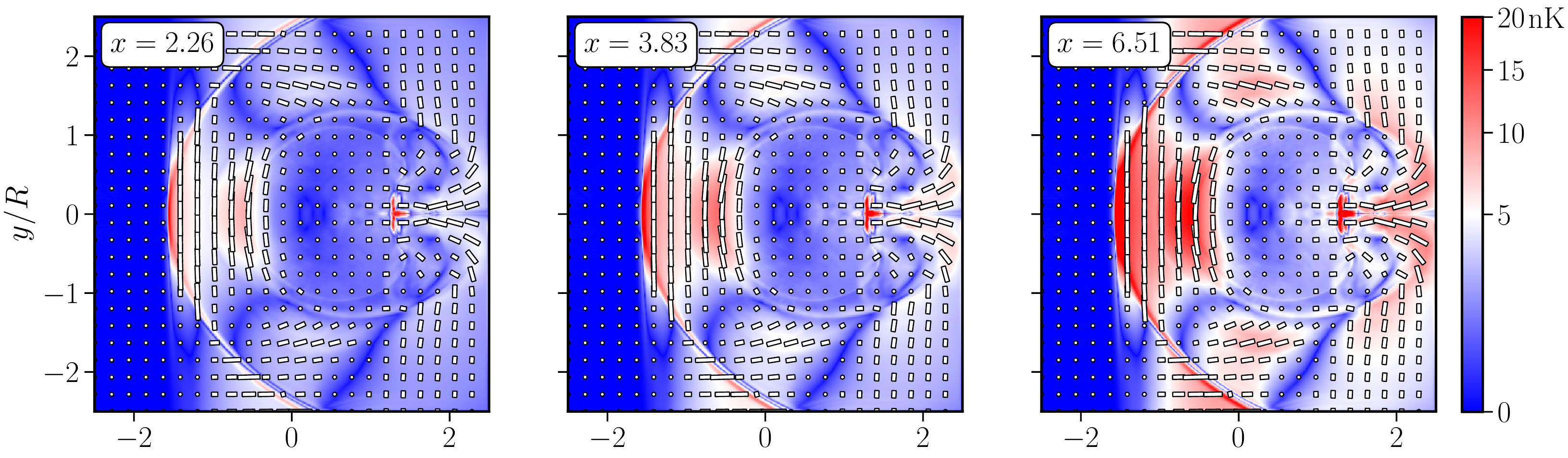}
\centerline{\labelstyle  Pressure-anisotropy-induced polarization, random initial magnetic field}
\includegraphics[width=0.9\textwidth]{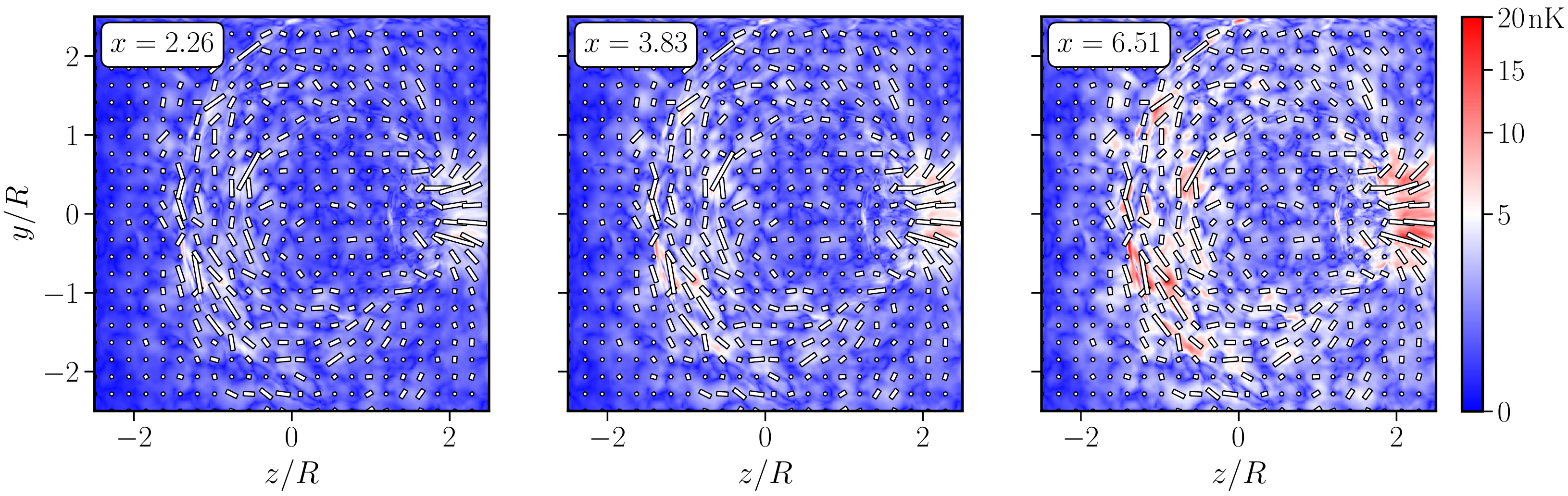}
%
\caption{ Various sources of CMB polarization in the direction of a galaxy cluster containing a (super)sonic cold front. The colour shows the amplitude of the signal (integrated along the line of sight over the computational domain), while the bars indicate the orientation of the polarization plane. Left panels correspond to frequencies around $ x=h\nu/kT_{cmb}=2.26$, middle panels to $x=3.83$, right panels to $x=6.51$. The top row shows the contribution of the kSZ-induced polarization, the row second from the top contribution of the $ \tau^2$ (i.e., induced by second scatterings) polarization, the row second from the bottom the contribution of the pressure-anisotropy-induced polarization in the case of a uniform initial magnetic field, the bottom row the contribution of the pressure anisotropy-induced polarization in the case of a random initial magnetic field.}
\label{f:szimage}
\end{figure*}

%% file: figure2.tex
\begin{figure*}
%
\centerline{\labelstyle Pressure anisotropy(uniform initial magnetic field)+ kSZ+$\tau^2$ polarization}
\includegraphics[width=0.9\textwidth]{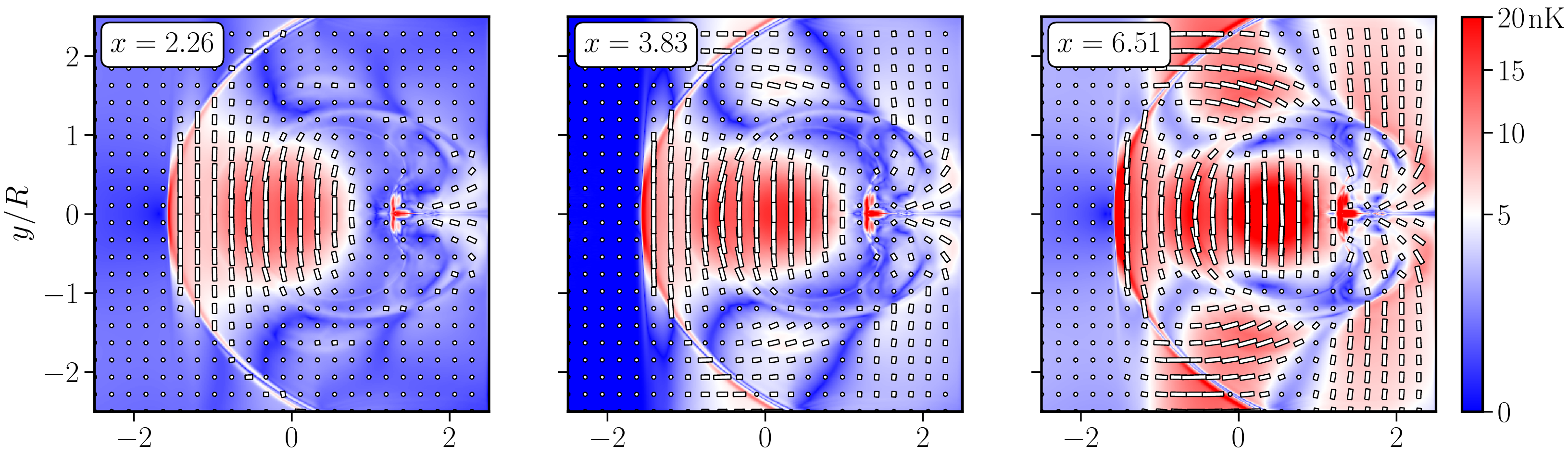}
\centerline{\labelstyle Pressure anisotropy(random initial magnetic field)+ kSZ+$\tau^2$ polarization}

\includegraphics[width=0.9\textwidth]{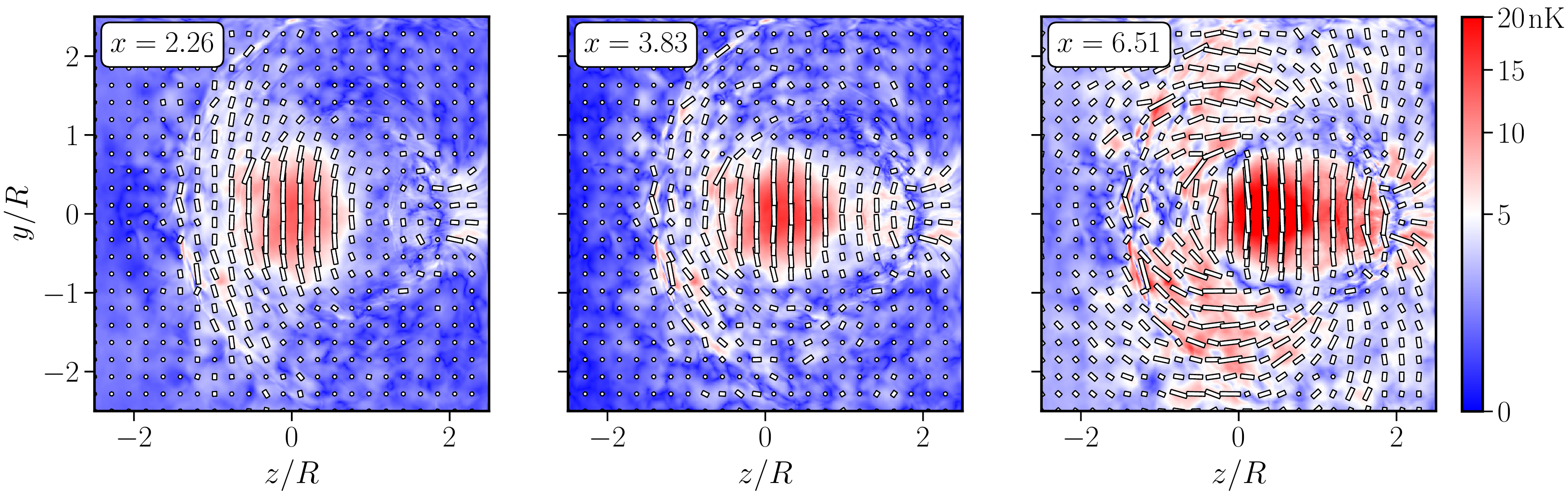}
\caption{The same as Figure \ref{f:szimage} but for the total predicted polarization signal for the uniform (upper row) and random (lower row) magnetic initial field configurations.}
\label{f:total}
\end{figure*}